\documentclass[10pt, conference, compsocconf]{IEEEtran}
\usepackage[utf8]{inputenc}
\usepackage[T1]{fontenc}
\usepackage[table]{xcolor}
\usepackage{color,soul}
\usepackage{graphicx,tabularx,url,amsmath,amssymb,tikz,float,siunitx,soul,verbatim,dsfont,pgfplots,subcaption,subfloat}
\usepackage[lined,ruled,linesnumbered]{algorithm2e}
\usepackage[pdftex, %
 			pdftitle={title}, %
 			pdfcenterwindow=true, %
 			pdfstartview=Fit, %
 			pdfdisplaydoctitle=true, %
 			pdfauthor={authors}, %
 			pdfpagemode=UseNone,
 			bookmarksopen=true,
 			breaklinks=false,
 			bookmarksnumbered=true,
 			colorlinks=true,
 			urlcolor=blue,
 			anchorcolor=black,
 			citecolor=black,
 			linkcolor=black]{}

\usepackage[textsize=scriptsize]{todonotes}
\newcommand\blfootnote[1]{%
  \begingroup
  \renewcommand\thefootnote{}\footnote{#1}%
  \addtocounter{footnote}{-1}%
  \endgroup
}

\makeatletter
\def\footnoterule{\kern-3\p@
  \hrule \@width 2in \kern 2.6\p@} 
\makeatother

\begin{document}

\title{Unsupervised Learning for Trustworthy IoT}
\author{
	\IEEEauthorblockN{Nikhil Banerjee, Thanassis Giannetsos, Emmanouil Panaousis, Clive Cheong Took}
	\IEEEauthorblockA{Department of Computer Science, University of Surrey, United Kingdom
    }
	}
\IEEEpubid{\copyright~2018 IEEE}

\maketitle
\begin{abstract}

The advancement of Internet-of-Things (IoT) edge devices with various types of sensors enables us to harness diverse information with Mobile Crowd-Sensing applications (MCS). This highly dynamic setting entails the collection of ubiquitous data traces, originating from sensors carried by people, introducing new information security challenges; one of them being the preservation of data trustworthiness. What is needed in these settings is the timely analysis of these large datasets to produce accurate insights on the correctness of user reports. Existing data mining and other artificial intelligence methods are the most popular to gain hidden insights from IoT data, albeit with many challenges. In this paper, we first model the cyber trustworthiness of MCS reports in the presence of intelligent and colluding adversaries. We then rigorously assess, using real IoT datasets, the effectiveness and accuracy of well-known data mining algorithms when employed towards IoT security and privacy. By taking into account the spatio-temporal changes of the underlying phenomena, we demonstrate how concept drifts can masquerade the existence of attackers and their impact on the accuracy of both the clustering and classification processes. Our initial set of results clearly show that these unsupervised learning algorithms are prone to adversarial infection, thus, magnifying the need for further research in the field by leveraging a mix of advanced machine learning models and mathematical optimization techniques.
\blfootnote{2018 IEEE. Personal use of this material is permitted. Permission from IEEE must be obtained for all other uses, in any current or future media, including
reprinting/republishing this material for advertising or promotional purposes, creating new
collective works, for resale or redistribution to servers or lists, or reuse of any copyrighted
component of this work in other works.}
\blfootnote{This research work is based upon the concept of the SPEAR project No.~787011 funded by the European Commission, within the H2020 initiative.}

\end{abstract}


\vspace{-0.2cm}
\section{Introduction and Background}
\label{sec:introduction}

During the last few years, the area of Internet-of-Things (IoT) has met a great development and has the potential to offer a new understanding of our environment that will lead to innovative applications with tangible positive impact on users' experience. This new paradigm leverages the proliferation of modern sensing-capable devices to build a wide-scale information collection network that can provide insights for practically, \emph{anything}, from \emph{anywhere} and at \emph{anytime}.
In order to provide concrete implementations for such complex environments, many challenges have to be overcome with \emph{security} and \emph{privacy} being critical pillars~\cite{giannetsos2014trustworthy}: especially in the context of safety applications where critical decisions are based on information collected by users regarding their status or surrounding events. For instance, in the civil protection space, this human-as-a-sensor paradigm is used to harvest information that enhances situational awareness~\cite{ballesteros2012safe}. 

In the past, extensive research has been conducted towards ``\emph{protecting the users from the system}'': building secure and accountable IoT architectures that can safe-guard user privacy while supporting user incentive mechanisms. Plethora of research efforts~\cite{shin2011anonysense, gisdakis2014sppear} have leveraged advanced cryptographic primitives (e.g., pseudonyms, group signatures, etc.) for protecting users' data from unauthorized access and preventing potential leak of personal identifiable information. However, the question, of how to ``\emph{protect the system from the users}'' in assessing the trustworthiness of contributed data so that strong guarantees can be provided towards the accuracy and correctness of the system output remains still open~\cite{gisdakis2015shield}.

In the machine learning community, security problems have already been addressed in the form of adversarial machine learning. For instance, novelty detection has been addressed to detect anomaly in acoustic data \cite{principi2017acoustic}. Game theory has also been exploited in the design of convolutional neural networks to detect image tampering \cite{chivukula2017adversarial}. Concept drift, which is a common phenomenon in IoT data, has also been considered in security problems such as in feature extraction \cite{cavalcante2016fedd} and fraud detection \cite{frauddetectionconceptdrift15}. Yet, most security/adversarial machine learning is based on the assumption that training data are readily available. As such, there are only a few works on unsupervised learning for IoT data such as anomaly detection \cite{ahmad2017unsupervised}. Our work addresses this shortcoming in the context of trustworthiness within IoT.

The trustworthiness of collected information is typically studied in relation to the trustworthiness of the human sensors which raises important concerns on the \emph{content integrity}. Data are not necessarily originated from trustworthy sources (e.g., sensors deployed and managed by authorities) but from contributions of any user volunteers that posses a sensing-capable device. This desired openness of IoT systems, such as Mobile Crowd-Sensing (MCS)~\cite{giannetsos2014trustworthy}, renders them vulnerable to malicious users that can \emph{pollute} the data collection process, thus, manipulating the system output \cite{rahman2017assessing}. A major challenge in these settings is the timely analysis of large amounts of data to produce highly reliable and accurate insights and decisions on the correctness of incoming user reports. Unfortunately, this is not straightforward especially in the presence of intelligent and colluding adversaries trying to manipulate the system's perception of the phenomenon. Data mining and other artificial intelligence methods are among the top methods to gain hidden insights from IoT data, albeit with many challenges.

Although assuring data trustworthiness is a classical security problem, especially in the context of sensor networks, existing approaches based on traditional unsupervised learning techniques will not suffice as explained next. For the aforementioned MCS systems, data types can range from simple environmental monitoring to more complex data describing dynamic and uncertain phenomena continuously evolving over space and time (e.g., traffic information systems). Therefore, well-studied techniques~\cite{dua2009towards} for responding to content integrity violations such as integrity model checking (requires historical data, which may not be available), reputation rankings (vulnerable to collusion), data aggregation (cannot provide high level of granularity) and independent human comparisons for data tagging (not feasible for continuous data streams), cannot be easily applied to IoT datasets. What is required is a \emph{set of mechanisms for assuring user-contributed information without prior statistical description of the data to be collected}. Towards this direction, a variety of works focus on building systems capable of handling available (contradicting) evidence, classifying efficiently incoming reports and effectively separating and rejecting those that are faulty~\cite{gisdakis2015shield}. However, their focus is on practicality aspects and their applicability to complex IoT datasets; no special attention has been given to the efficacy and accuracy of the internally employed machine learning algorithms.

\emph{\textbf{Motivation \& Contributions}:}~The aim of this research is to introduce and highlight this cyber-trustworthiness aspect in IoT environments and explore \emph{how} to leverage conventional data mining algorithms for preempting adversarial behaviour. To this end, we focus on MCS applications as a use case where sensing attacks are prominent for users to save their sensing costs and avoid privacy leakage, and we assume the employment of state-of-the-art data verification frameworks, such as the one described in~\cite{gisdakis2015shield}. The main contribution of this work is to provide a comprehensive analysis on the applicability of various clustering and classification techniques for identifying descriptive and predictive models to classify non-faulty and faulty user incoming reports. More specifically, we (\emph{i}) provide a rigorous assessment of the effectiveness, efficiency and accuracy of several well-known data mining algorithms (Support Vector Machine (SVM), Naive Bayes (NB), Random Forest (RF), and neural Networks (NNs)) when employed towards enhancing IoT security and privacy (deviating from existing works that solely focus on performance based on one-shot solutions and adversarial-free models~\cite{chen2015data, feng2016survey}, conditions that fail to hols in safety-critical applications), (\emph{ii}) evaluate the impact (level of distortion) to the system output in the presence of strong, colluding adversaries, and (\emph{iii}) demonstrate how the spatio-temporal changes on the underlying phenomena (i.e., concept drift) can masquerade the existence of attackers and affect the accuracy of both the clustering and classification processes. We extensively evaluate the performance of the unsupervised learning techniques, under various scenarios, employing both real and synthetic datasets. 

In the rest of this paper, we first describe the problem statement and define the system and adversarial models (Sec.~\ref{sec:problem_statement}). We then provide an overview of MCS and the details of the employed data verification framework focusing on the internal clustering and classification processes (Sec.~\ref{sec:mcs}). Sec.~\ref{sec:setup} outlines the experimental setup used to evaluate our core unsupervised learning algorithms, with results presented in Sec.~\ref{sec:performance}. Based on our findings, we posit open issues and challenges and discuss possible ways to address them in Sec.~\ref{sec:discussion} before we conclude in Sec.~\ref{sec:conclusion}; so that machine learning can offer enhanced IoT security and privacy capabilities that can further accelerate data-driven insights and knowledge acquisition.

\begin{figure*}
    \centering
    \includegraphics[width=0.7\textwidth]{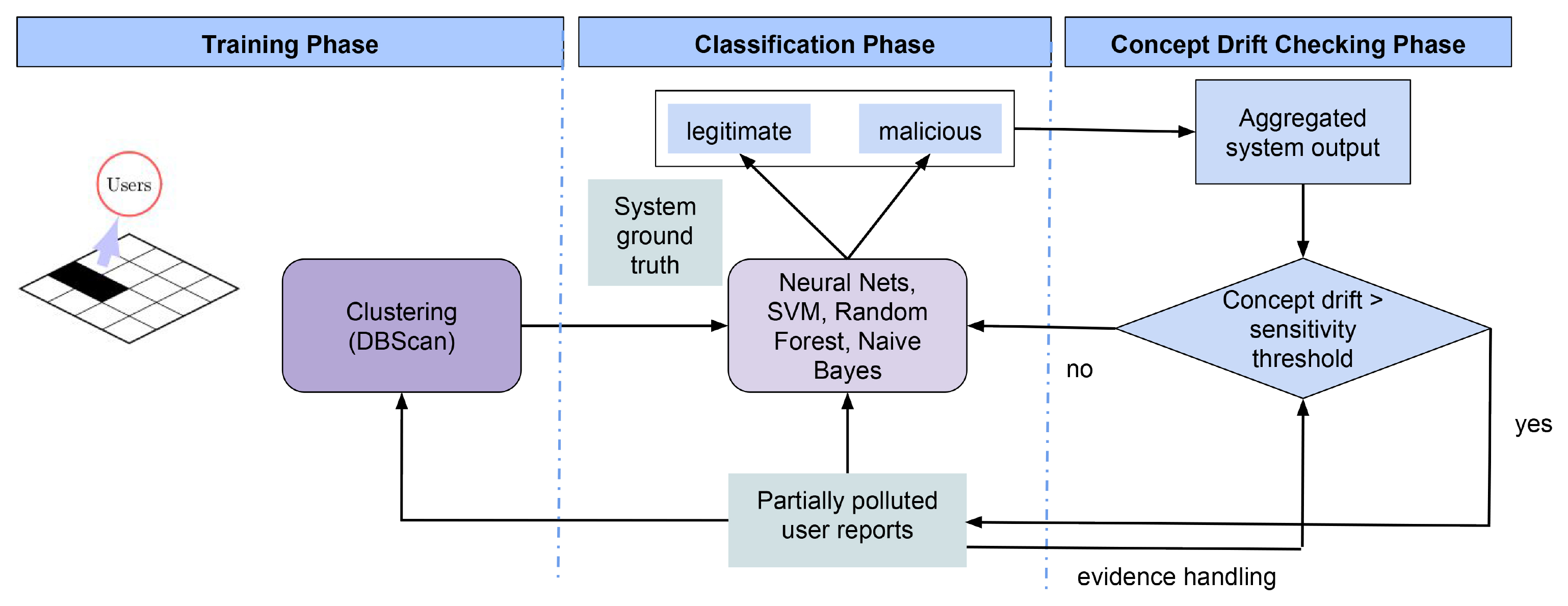}
    \caption{MCS Data Verification.}
    \label{fig_mcs}
    \vspace{-0.4cm}
\end{figure*}

\section{Cyber Trustworthiness of IoT MCS Reports}
\label{sec:problem_statement}

\emph{\textbf{System Model}:}~We consider the employment of a Mobile Crowd-Sensing~\cite{giannetsos2014trustworthy} platform (Sec.~\ref{sec:mcs}) consisting of a large set of users operating mobile devices equipped with embedded sensors (e.g., inertial and proximity sensors, cameras, microphones) and navigation modules. User mobile devices collect and report sensory data, based on a specific \emph{sensing task} published by the MCS infrastructure (for a specific location), and report back their raw data over any available network.

Each user submits to the infrastructure a stream of measurements on the sensed phenomenon over a time interval, $t$, specified in the sensing task description. We assume the existence of a secure and privacy-preserving architecture, like the ones presented in~\cite{shin2011anonysense, gisdakis2014sppear}\footnote{We omit further cryptographic protection specifics as these have been addressed in the literature and we refer the reader to the cited papers.}, for guaranteeing the secure communications and enhancing the privacy posture of the user devices in order to avoid sensitive information leakage and/or extensive user-profiling. User data are submitted in successive \emph{reports}, each with $n$ measurements, $v_i$ where $i\in\{1, 2,..., n\}$, corresponding to a device location, $loc$:

\begin{center}
	$r_i = \left \{ [v_1,v_2,v_3,...,v_n] \left | \right | t \left | \right | loc \left | \right |  \sigma_{PrvKey} \left | \right | C \right \}$
\end{center}

$\sigma_{PrvKey} $ is a digital signature with some private key, with the corresponding public key included in the certificate $C$.

\emph{\textbf{Threat Model}:}~The aim of the adversarial agents is to mislead the IoT system towards desired malicious measurement values $T$. To this end, this can be achieved by maximizing the following constrained function, that is,
\begin{equation}
\max f[x^{t} -x^{(a)}(\tau)]
\textrm{~~~~~~s.t.~~~} f\leq T
\end{equation}
where $f$ denotes a utility function of the adversarial agent, $x^{t}$ represents the actual ground truth measurement value and $x^{(a)}$ denotes the malicious measurement sample. The function can be linear or nonlinear and be more complex in formulation as in \cite{dalvi2004adversarial}. Our contribution is to model for the ``drift'' strategy of the adversarial agent similar to concept drift in machine learning.
The $i$-th agent, thus, introduces a misleading measurement sample at time $\tau$, which can be modelled as
\begin{equation}
\small
x_i^{(a)}(\tau)= \min(x_i^{t}+ \tau\delta+\eta_i(\tau), T),~~\forall~i=\{1,...,N_a\}, \label{eq: adversarial agent}
\end{equation}
such that
\begin{eqnarray}
\tau=0&&x_i^{(a)} =x_i^{t}+\eta_i(0)\approx x_i^{t}\nonumber\\
\tau=1&&x_i^{(a)} = \min(x_i^{t}+ \delta+\eta_i(1), T)\nonumber\\
\tau=2&&x_i^{(a)} = \min(x_i^{t}+ 2\delta+\eta_i(2), T)\nonumber\\
\hdots&&\nonumber\\
\tau\rightarrow\infty && x^{(a)}= T
\end{eqnarray}

where $\delta$ is a small positive number used to slowly drift from the `legitimate' measurement of the adversarial agent's device to the targeted final malicious measurement value $T$, and $\eta_i(\tau)$ is a small noise component modelled by Gaussian distribution at time instant $\tau$. The IoT system provides $N$ samples $\{x_i\}$ made up of legitimate measurements $\{x^{t}_i\}$ and malicious measurements $\{x^{(a)}_i\}$ generated by $N_a$ adversarial agents.
To maximize the trustworthiness of the IoT system, the overarching objective is to detect the legitimate measurements $\{x^{t}_i\}$ from the malicious measurements $\{x^{(a)}_i\}$. In MCS, this problem is further compounded by the fact that we do not have access to labelled or training data $\{x^{t}_i\}$. Thus, the classification problem is inherently an unsupervised problem. To detect the legitimate measurements, the problem can be thus simplified into two sub-problems: clustering of $N$ samples $\{x_i\}$ to artificially create labelled data of two categories, and then the classification of the malicious measurements from the legitimate measurements by exploiting the labelled data.
\\\noindent \textbf{Remark$\mathbf{\#1}$}:~Clustering inherently performs classification, thus, using a classifier may seem redundant. However, it is crucial to use both. It is not practical to run a clustering algorithm whenever a new sample is received by the IoT system, but it is computationally more affordable to run a trained classifier, especially since IoT requires lightweight applications.
\\\noindent \textbf{Remark$\mathbf{\#2}$}:~The model used by adversarial agents in (\ref{eq: adversarial agent}) show that they are mimicking concept drift as a phenomenon. For example, it is not uncommon for temperature to drift from 30 degrees during the day to 16 degrees during the night. However, the model in (\ref{eq: adversarial agent}) is unlikely to be drifting at the same rate of that of a natural phenomenon such as temperature. Thus, regular clustering on new samples for classification may lead to false detection. On one hand, updating the labels too often by clustering leads to biasing the IoT system towards the malicious measurements. On the other hand, we need to account for the concept drift of natural phenomenon to properly update the labels of the data for the classifier to be more robust in a dynamic environment.

\section{Mobile Crowd-Sensing \& Data Verification}
\label{sec:mcs}

The advancement of smart-phones with various type of sensors enabled us to harness diverse information with Mobile Crowd-Sensing (MCS) applications. In particular, an MCS platform or server recruits mobile users to monitor the surrounding features and offers crowd-sensing applications spanning from urban sensing~\cite{miluzzo2011tapping} and network and traffic monitoring~\cite{hull2006cartel, whitefield2017privacy}. 

The features of interest (to be sensed) are described in a \emph{sensing task} description that is essentially a data collection campaign distributed to recruited users~\cite{giannetsos2014trustworthy}. The \emph{area of interest} of a sensing task is the locality within which participating users must contribute data. The area of interest can be defined either explicitly (e.g., coordinates
forming polygons on maps) or implicitly (through annotated geographic areas). In most cases, the area of interest is divided into spatial units: homogeneous, with respect to the sensed phenomenon, areas. This is mainly due to the following reason: In an MCS environment, there are a wide variety of devices by different users and it is not possible to retrieve ground truth from the users. Thus, there is no a priory knowledge of what makes reports faulty or non-faulty; something that needs to be extracted by analyzing data objects without consulting a known class model. The sensed phenomenon has temporal but not significant spatial variations within a spatial unit, thus, allowing the exploration of the spatial characteristics (of the incoming data) towards \emph{reasoning} on the actual value of the sensed phenomenon.


As aforementioned, we assume the existence of a data verification framework, similar to the one presented in~\cite{gisdakis2015shield}, for assessing the data trustworthiness of user incoming reports. In what follows, we take take a similar approach, as illustrated in Fig. \ref{fig_mcs}. We consider 3 phases as follows:

\noindent \textbf{Training phase}: During the training phase, legitimate user reports are classified as \textit{legitimate} or \textit{malicious} corresponding to non-faulty and faulty ones. Based on its design, the system does not know with \emph{certainty} whether a report is legitimate or malicious. Incoming user reports are classified as evidence rather than raw data (based on the Dempster-Shaffer Theory). 
 
\noindent \textbf{Classification phase}: In this phase, the output of the training is entered into the employed classifiers (i.e., NB, SVM, RF and NNs). These learning algorithms are used to classify the users' reports received after the end of the training phase.

\noindent \textbf{Concept drift checking phase}: The statistical properties of the sensed data may change over time causing what we call as \textit{concept drifts}. These can deteriorate the performance of the classification and they impose the need for quick adaptation so that MCS data trustworthiness is maintained. 

In the investigated problem, we assume that there is no ground-truth to train the MCS application. We, thus, need to take a hybrid clustering-classification approach. Although clustering by itself can be used for classification, it is not as efficient as classification when it comes to classify every single incoming report for each spatial unit. As a result of this, we consider that after clustering occurs there is a period $T$ during which only classification is executed on the incoming reports, which have been partially (i.e., less than 50\%) infected with adversarial samples. The period parameter $T$, before re-clustering is triggered, is determined by a \emph{sensitivity threshold}, $\theta$, which is compared against the current perception of the system for the sensed phenomenon (current value), as described above. If a conflict between the already computed ground-truth and the incoming user reports exceeds, $\theta$, the system switches from the classification to the clustering phase in order to re-train the underlying descriptive data model.

Unfortunately, we cannot assume that the re-training will consist of adversarial-free samples because we have assumed that data is legitimate on its entirety only at the beginning of the system setup (i.e., during the first time clustering occurs). An important challenge is the computation of $\theta$ so that an optimal balance between performance and accuracy is achieved. In other words, $\theta$ must be defined such as \emph{concept drift} and \textit{attack drift} (Sec.~\ref{sec:performance}) are optimally distinguished, where the term attack drift refers to a concept drift that has been imposed by the adversarial samples. To give a better intuition of the role of $\theta$, note that $T$ decreases with $\theta$, i.e., the system is more sensitive for higher values of $\theta$ and the concept drift reaches $\theta$ faster, thus, triggering re-clustering. 

Based on the above discussion, this paper is a step towards investigating optimal behaviours of systems that use data mining algorithms to perform MCS capabilities. Among other things, in Sec.\ref{sec:performance}, we investigate the impact that attacks have into the re-clustering process and how this affects the subsequent classification. More precisely, we derive various results that show the accuracy of classification after the system has been re-trained with various levels of adversarial samples. 


\begin{figure*}[!ht]
  \centering
  \begin{minipage}[htp]{0.45\textwidth}
    \centering
    \includegraphics[width=1\textwidth]{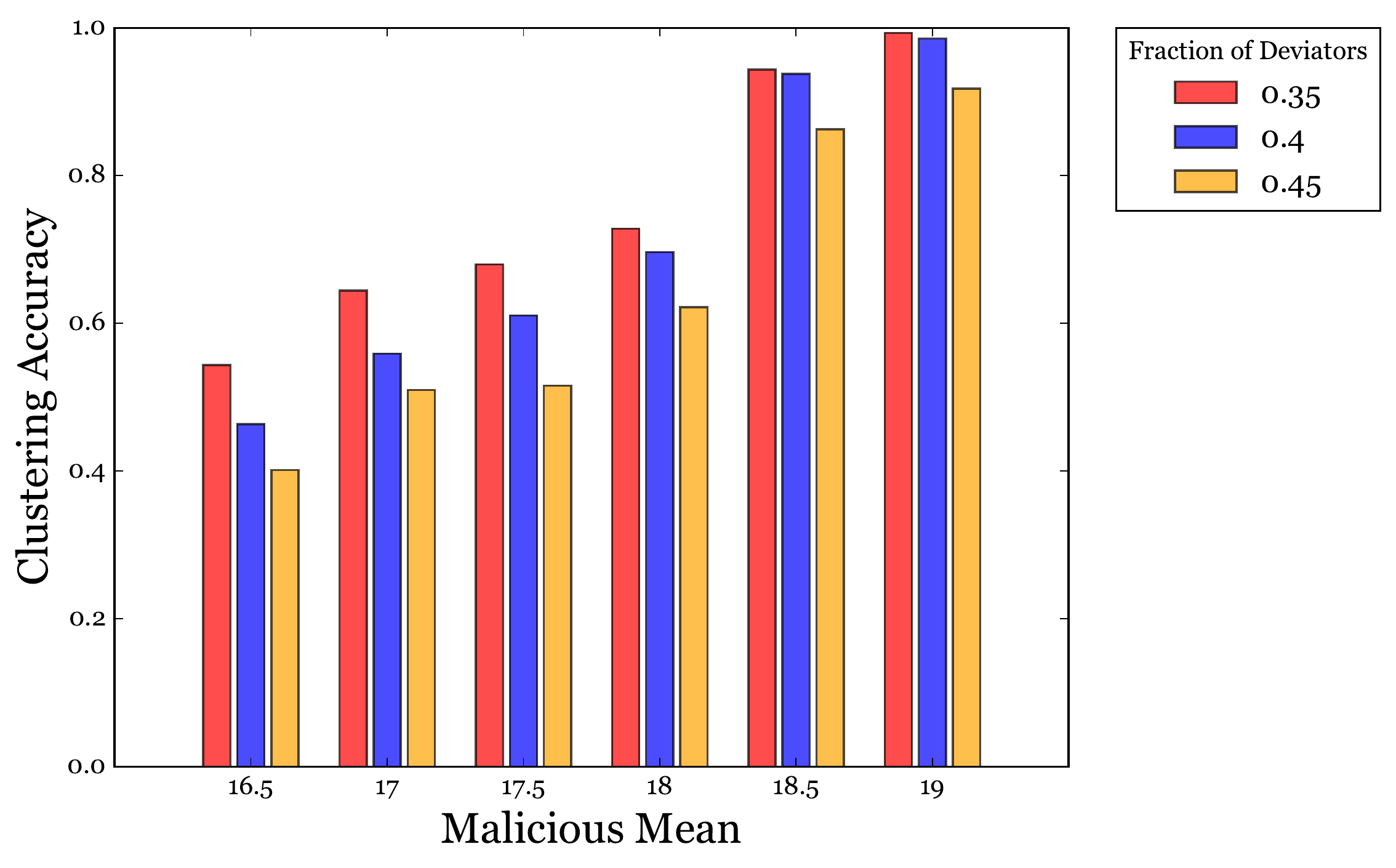}
    \vspace{-0.6cm}%
    \caption*{\footnotesize (a) Clustering precision for different fractions of deviating\\ users and varying distribution mean value, $\mu$.}
  \end{minipage}
    \centering
  \begin{minipage}[htp]{0.435\textwidth}
    \centering
    \includegraphics[width=1\textwidth]{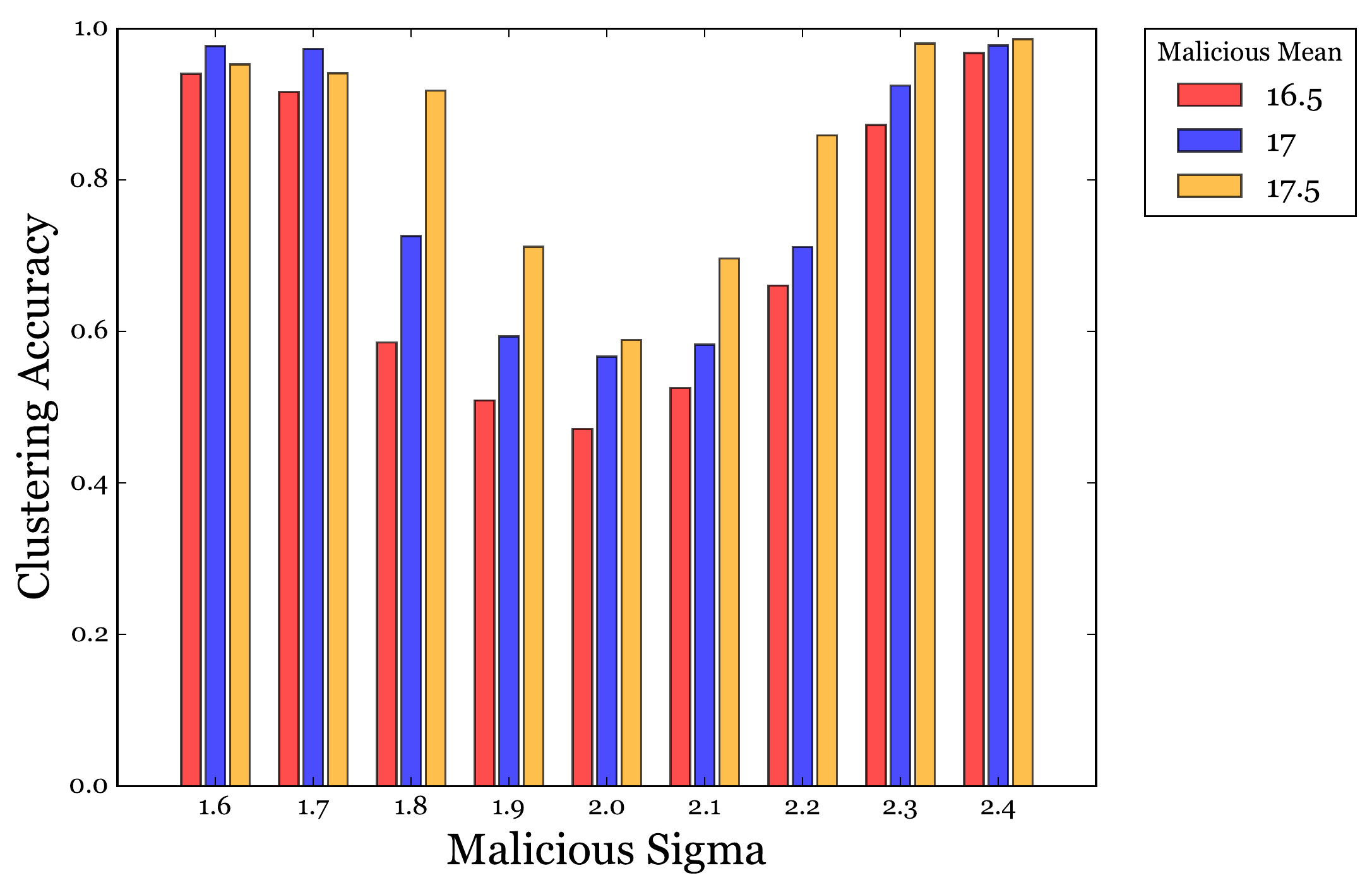}
    \vspace{-0.6cm}%
    \caption*{\footnotesize (b) Fixed fraction of deviating users (35\%) and varying\\ distribution standard deviation, $\sigma$.}
  \end{minipage}
  \caption{Performance analysis of the DBSCAN clustering algorithm.}
    \label{fig_clustering}
    \vspace{-0.4cm}
\end{figure*}

\vspace{-0.1cm}
\section{Experimental Setup}
\label{sec:setup}

The main focus of this work is to provide a comprehensive analysis and evaluation of various classification techniques used for assessing user incoming reports in each MCS region (Sec.~\ref{sec:mcs}); i.e., their characterization as \emph{legitimate} and \emph{malicious}. To this end, we provide strong empirical evidence on the performance of the most prominent algorithms leveraging both real-world and synthetic datasets. In order to obtain more effective and accurate results, four different classification algorithms are being explored, namely Neural Networks (NNs), Support Vector Machine (SVMs), Random Forest (RF) and Naive Bayes (NB)~\cite{mavrogiorgou2017comparative}, whose performance is being compared in terms of accuracy under the presence of strong, colluding adversaries. We opted in for using this sub-set of well-known data mining algorithms based on their efficiency to work within complex and noisy domains as is usually the case in IoT~\cite{abdar2017performance, li2015detection, kim2016collaborative}.

\emph{\textbf{Datasets}:}~We evaluate the system under various scenarios by employing both real-world and synthetic datasets. \emph{Real-world datasets} provide us with a good understanding of a classifier's performance in real case scenarios, i.e., deployed sensors measuring noisy physical phenomena. \emph{Synthetic datasets} allow us to estimate the impact of various system parameters (e.g., the distribution that legitimate reports follow) for different adversarial presence (i.e., number of malicious users ) and strategies (i.e., the distributions that malicious users employ). 

The real dataset has been originated from urban sensing applications~\cite{mendez2011p}, whereas the synthetic dataset relates to an MCS traffic monitoring. Regardless of the underlying set, we inject faulty adversarial reports and assess the classifier's ability to accurately achieve a truthful view of the underlying phenomenon. In our experiments, we used the Strata Clara dataset, from the Data Sensing Lab~\cite{dataset}, which lies within the domain of environmental monitoring~\cite{miluzzo2011tapping}. This dataset contains raw measurements of different physical phenomena (i.e., humidity, sound and temperature), from 40 sensors deployed at the Strata Clara convention center in the United States in 2013. The normal distributions of the monitored phenomena are: $(\mu, \sigma) = \{(31, 5) | (3, 2) | (21, 1.3)\}$, respectively.

The \emph{synthetic dataset} emulates a traffic monitoring MCS task~\cite{gisdakis2015secure}. It has been generated by drivers' smartphones that are reporting their location and velocity. We consider 250 users and simulate urban road links along with their traffic conditions by generating ``actual'' location traces for each vehicle (i.e., smartphone) using the Simulation of Urban MObility (SUMO) traffic simulator \cite{fernandes2010platooning}. 

\emph{\textbf{Adversarial Behaviour}:}~The overall goal of the framework is to infer the actual value of the sensed phenomenon in presence of adversarial users who generate malicious reports to set the system perception to a faked value. We assume that adversaries collaborate to attack the data collection process. The collaboration is achieved by having injected malicious reports drawn from the same normal distributions, which are also different to those of the adversary-free datasets (the legitimate reports that determine the actual phenomenon). 

In our experiments, we measure the performance of the various classifiers in terms of their resistance to adversarial data infection. The adversarial choices determine the distortion adversaries try to impose on data trustworthiness~\cite{biggio2013evasion}. The chosen values for $(\mu, \sigma)$ determine the similarity (overlap) between the legitimate and adversarial distributions. Intuitively, adversarial detection decreases with this similarity because the malicious reports introduce values that are very near to the legitimate ones. To increase the probability to detect adversarial reports, which are entered to the machine learning model, we consider the following cases:

\noindent\textbf{Case I}:~Adversaries may cause significant distortion of the phenomenon's values by increasing the distance between $\mu$ of the adversarial distribution and $\mu$ of the legitimate distribution;

\noindent\textbf{Case II}:~Adversaries may maximize the system uncertainty by choosing a normal distribution with large $\sigma$;

\noindent\textbf{Case III}:~Adversaries may increase the system uncertainty about the true value of the phenomenon by selecting an adversarial distribution with $\mu$ equal to $\mu$ of the legitimate distribution but with significantly smaller $\sigma$.

\section{Results and Analysis}
\label{sec:performance}

The effects of the aforementioned classification algorithms are based on the following parameters, i.e, size of dataset, performance and accuracy. In what follows, we analyze the system's ability to identify and filter out faulty reports, i.e., the labeling of user reports as legitimate and malicious.

The performance metrics of interest used to evaluate and interpret the classifiers' results are the following: (\emph{i}) \emph{confusion matrix} that contains information about the classifications' results, (\emph{ii}) \emph{true positive rate} that depicts the percentage of correct predictions, (\emph{iii}) \emph{false positive rate} the reflects the proportion of instances classified in class $x$, but belong to a different class, along with all the instances that are not in class $x$, (\emph{iv}) \emph{recall} that depicts the proportion of instances that are correctly predicted as positive, and (\emph{v}) \emph{precision} that estimates the probability that a positive prediction is correct.

In each experiment, we provide the system with reports originating from both legitimate and malicious users (Sec.~\ref{sec:setup}). This dataset is partitioned into two sub-sets: a training set (TS) and an evaluation set (ES). Based on the TS, the clustering (training phase) takes place. Then, the ES is used to assess the performance of the supervised classification part. For each simulation we perform tenfold cross validation to avoid overfitting. We show results based on both real and synthetic datasets; due to space limitations, we do not repeat similar figures from both sets.

\emph{\textbf{Clustering accuracy}}:~Since the clustering is not efficient for continuous data streams (requires the execution of the clustering algorithm for each incoming report (Sec.~\ref{sec:mcs})), we refrain from evaluating different algorithms. We focus primarily on the impact the accuracy of the employed clustering (i.e., DBSCAN) has on the performance of the classifiers. 

Fig.~\ref{fig_clustering} examines the accuracy of the DBSCAN clustering algorithm for different fractions of deviating users and varying adversarial strategies (data distributions). We leverage the synthetic datasets generated from the emulated traffic sensing task. Legitimate users report their velocity from a normal distribution with $(\mu = 16, \sigma = 2)$. Our focus is on measuring the level of distortion that adversaries can incur on the system when trying to identify patterns from the collected data so as to gain a perception on the sensed phenomenon. Adversarial behavior can employ different strategies (Sec.~\ref{sec:setup}) based on whether they want to target the systems' \emph{certainty} or \emph{uncertainty} of the phenomenon; by crafting data distributions with selected $\mu$ and $\sigma$ values that correspond to a different percentage of overlapping regions (Sec.~\ref{sec:setup}).

Fig.~\ref{fig_clustering} (a) takes a closer look at DBSCAN's accuracy when adversaries follow a normal distribution with a fixed standard deviation, $(\sigma = 2)$, (equal to the one of the legitimate data distribution) and varying mean values, $\mu$. Here, we set the number of malicious users to be 35$\%$, 40$\%$ and 45$\%$, respectively. We see that DBSCAN achieves (almost) perfect clustering when the overlap between the distributions is relatively small (i.e., difference in mean values is higher than 15$\%$) and even for a large number of adversaries (45$\%$). When the overlap is around 10$\%$ (adversarial $\mu \in [17, 18]$), the clustering accuracy still remains relatively high ($\geq 60\%$). Only for high overlap percentages, the accuracy drops to close to 50$\%$ (or lower in the case of a high number of adversaries).

\begin{figure*}[!ht]
  \centering
  \begin{minipage}[htp]{0.24\textwidth}
    \centering
    \includegraphics[width=1\textwidth]{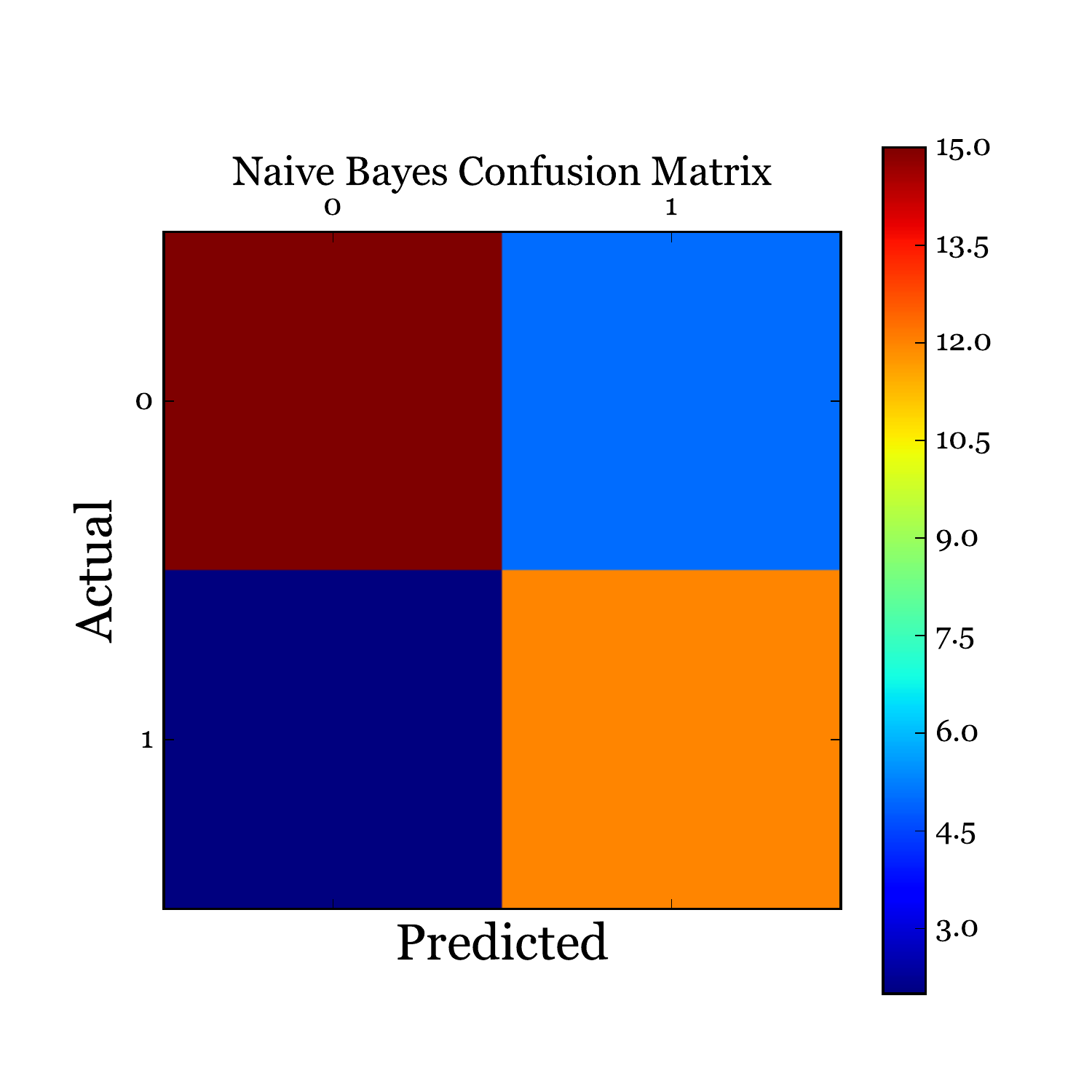}
    \vspace{-0.6cm}%
    \caption*{\footnotesize (a) Naive Bayes (NB).}
  \end{minipage}
    \centering
  \begin{minipage}[htp]{0.24\textwidth}
    \centering
    \includegraphics[width=1\textwidth]{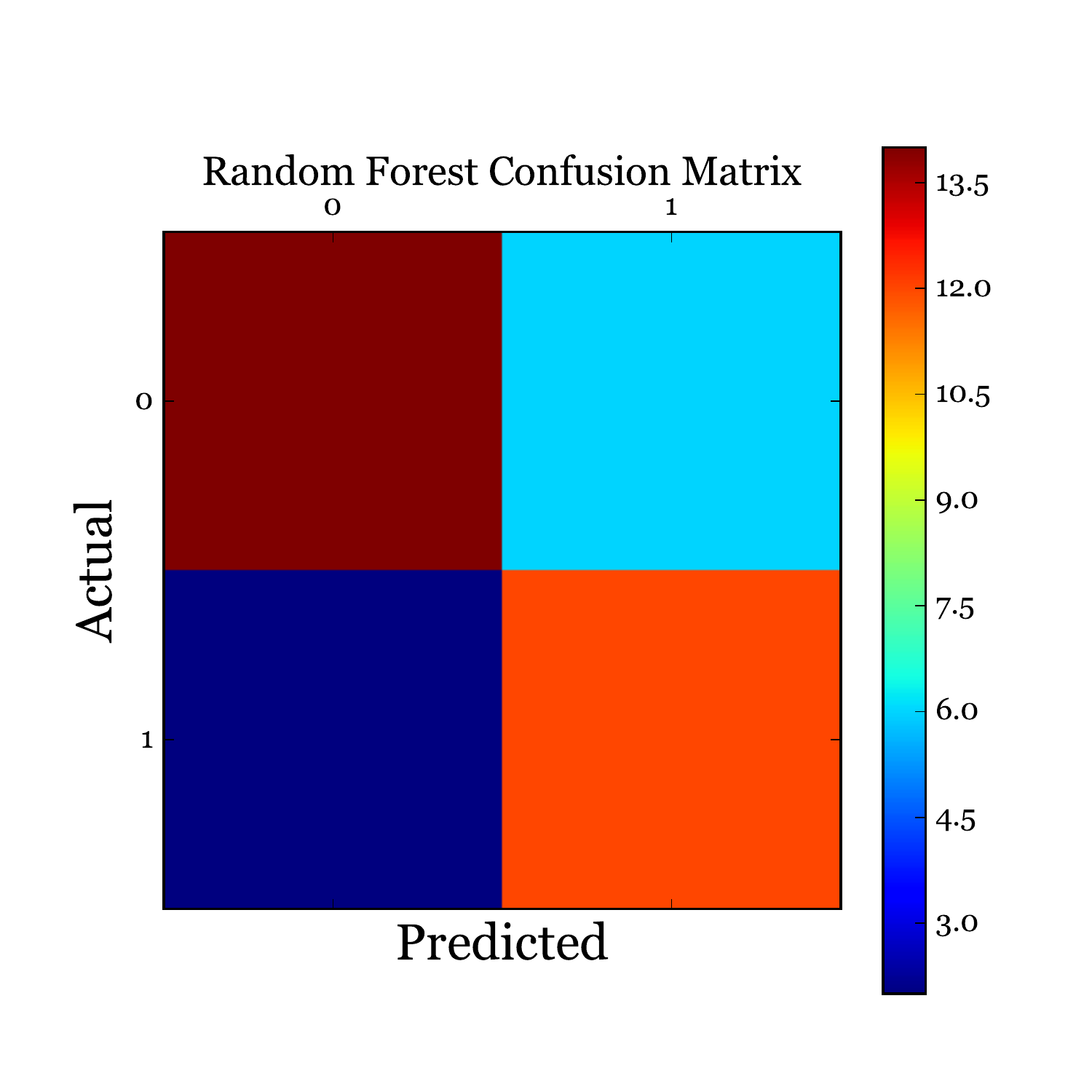}
    \vspace{-0.6cm}%
    \caption*{\footnotesize (b) Random Forest (RF).}
  \end{minipage}
  \begin{minipage}[htp]{0.24\textwidth}
    \centering
    \includegraphics[width=1\textwidth]{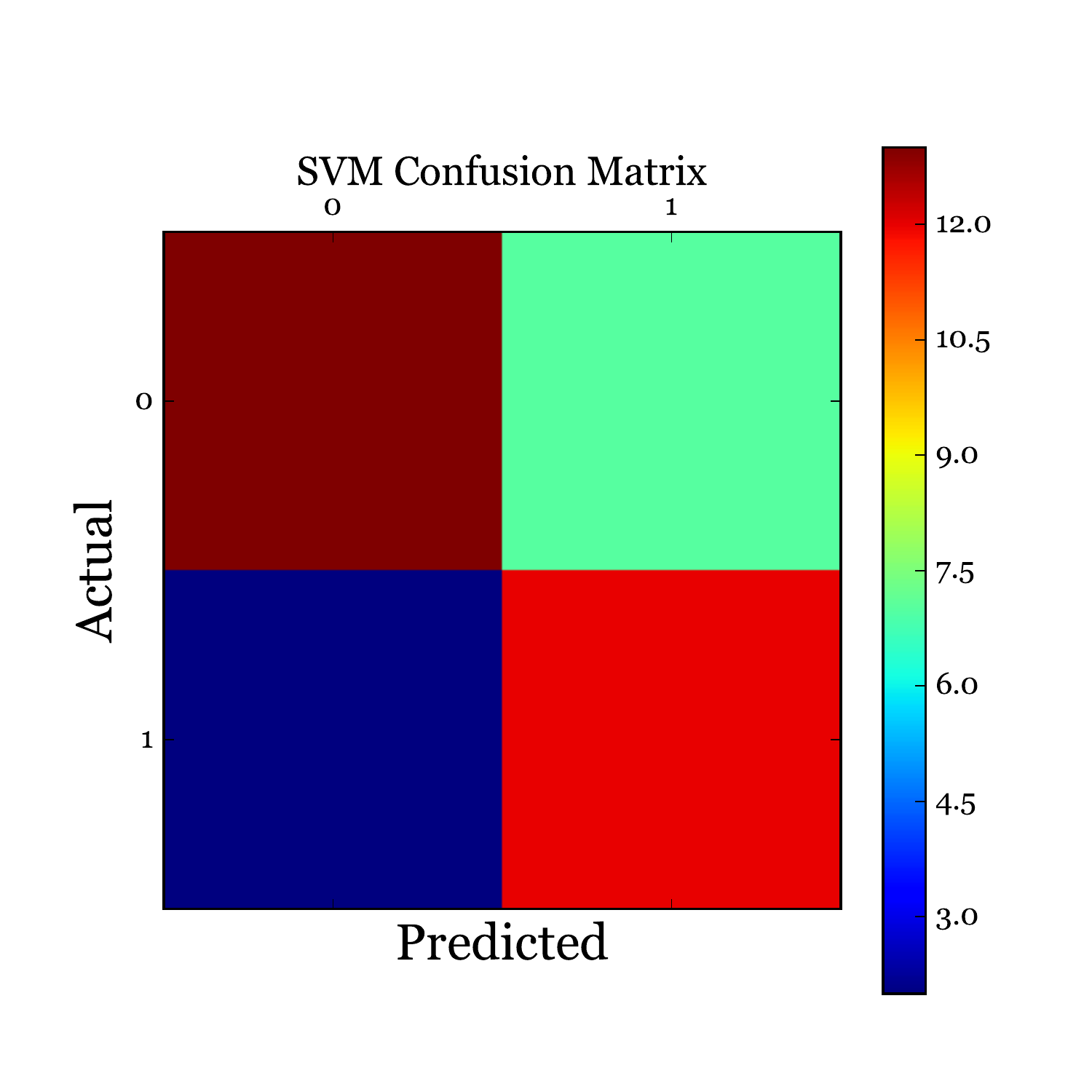}
    \vspace{-0.6cm}%
    \caption*{\footnotesize (c) Support Vector Machine (SVM).}
  \end{minipage}
  \begin{minipage}[htp]{0.24\textwidth}
    \centering
    \includegraphics[width=1\textwidth]{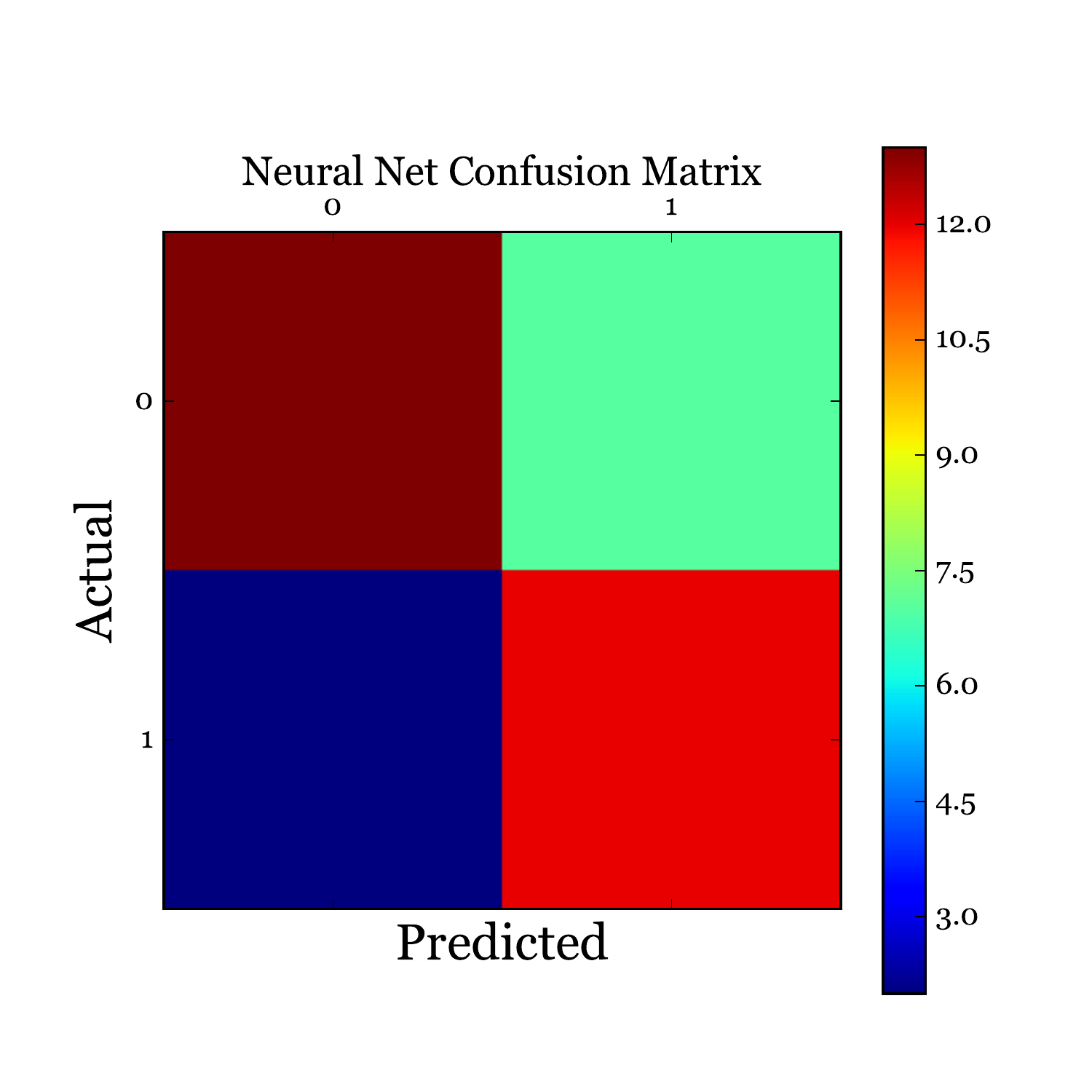}
    \vspace{-0.6cm}%
    \caption*{\footnotesize (d) Neural Networks (NNs).}
  \end{minipage}
  \caption{Confusion matrices for temperature $(\mu = 21, \sigma = 1.3)$ sensor and different classification algorithms.}
    \label{fig_confusion}
    \vspace{-0.4cm}
\end{figure*}

\noindent \textbf{Remark$\mathbf{\#3}$}: When the two distributions are almost identical, the clustering accuracy drops radically. However, the impact in this case is minimal since the adversaries do not significantly distort the system's output.

This is further demonstrated in Fig.~\ref{fig_clustering} (b) where we examine the impact of varying standard deviation values. Here, we fix the number of malicious users to be 40$\%$ and we vary the applied distributions (both $\mu$ and $\sigma$). We see that only in the case where the standard deviation values of both distributions are equal, the accuracy drops to ($\simeq 55\%$). In this case, the same $\sigma$ introduces ambiguity to the certainty of the system with regards to the sensing phenomenon, thus, making it difficult to decide on the correct descriptive data model.

Finally, for identical distributions, the clustering exhibits an average random behavior since malicious and legitimate reports do not differ at all (same as classifying based on a ``coin toss'').

\emph{\textbf{Classification accuracy}:}~Fig.~\ref{fig_confusion} depicts the accuracy of the employed classifiers for the SC dataset (temperature sensor) and for 40\% of deviating users following the same malicious distribution, $(\mu = 22, \sigma = 2)$. In this case, classification accuracy is represented in a structure termed \emph{confusion matrix}. Each column of the matrix shows the instances in a predicted class (1 for positive, i.e., legitimate and -1 for negative, i.e., malicious,
reports), while each row shows the instances in an actual class. A confusion matrix C is such that $C_{i, j}$ is equal to the number of observations known to be in $class_i$ but labeled, by the classifier, to be in $class_j$. Essentially, C shows the true positives (TPs), false positives (FPs), true negatives (TNs) and false negatives (FNs). The confusion matrices for NB, RF, SVM and NNs classifiers are shown in Fig. \ref{fig_confusion}. The diagonal elements show the number of correct classifications made for each class, and the off-diagonal elements indicate the errors.

\noindent \textbf{Remark$\mathbf{\#4}$}: The overall correctness of all employed classifiers remains high (Accuracy $\simeq 0.9$), despite the high number of malicious users and the high level of similarity between the malicious and legitimate data distributions. In almost all cases, the percentage of correctly classified samples is at least 85$\%$. Considering the high number of adversaries, this is partially due to the number of negative samples increasing in the evaluation set and, thus, mis-classifying one such sample has a smaller impact on the overall precision.

For instance, in the case of SVM (Accuracy $\simeq 0.9$), 17 out of 20 samples were classified positive correctly and all true negatives were identified without error (similar for NB and RF). In the case of NNs, the accuracy is slightly lower (Accuracy $\simeq 0.85$); 15 out of 20 samples were classified positive correctly whereas all true negatives were (again) identified without error. This minor drop in the accuracy may be due to the fact that NNs are more likely to overfit and can suffer from multiple local minima compared to SVMs. Overall, the true negative classification remains high in all cases and only the false positive rate shows a variation, depending on the internal way of operation of each classifier~\cite{mavrogiorgou2017comparative}. 

\begin{figure*}[!ht]
  \centering
  \begin{minipage}[htp]{0.45\textwidth}
    \centering
    \includegraphics[width=1\textwidth]{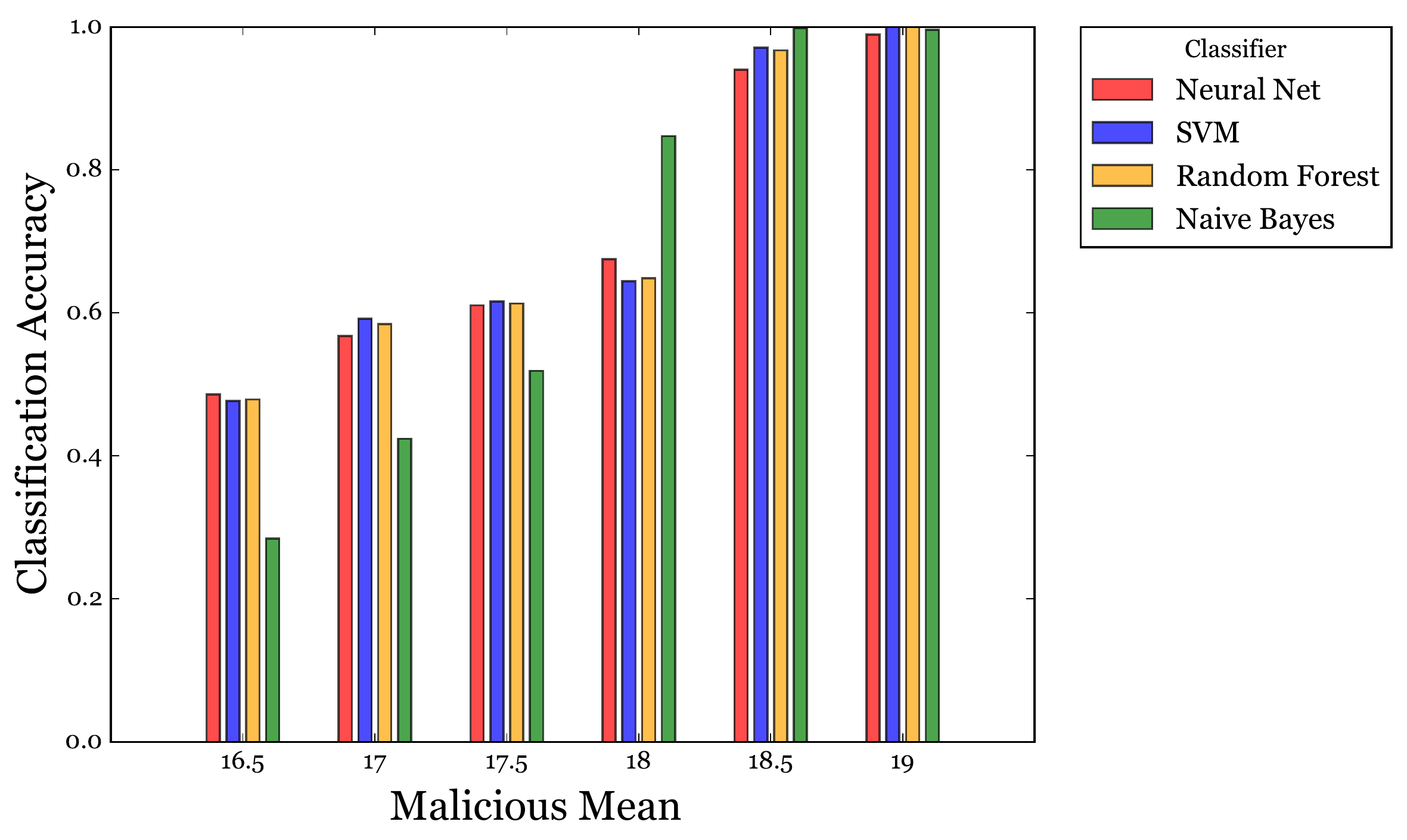}
    \vspace{-0.6cm}%
    \caption*{\footnotesize (a) (\emph{exploratory}) different ``normal'' malicious data distributions.}
  \end{minipage}
    \centering
  \begin{minipage}[htp]{0.45\textwidth}
    \centering
    \includegraphics[width=1\textwidth]{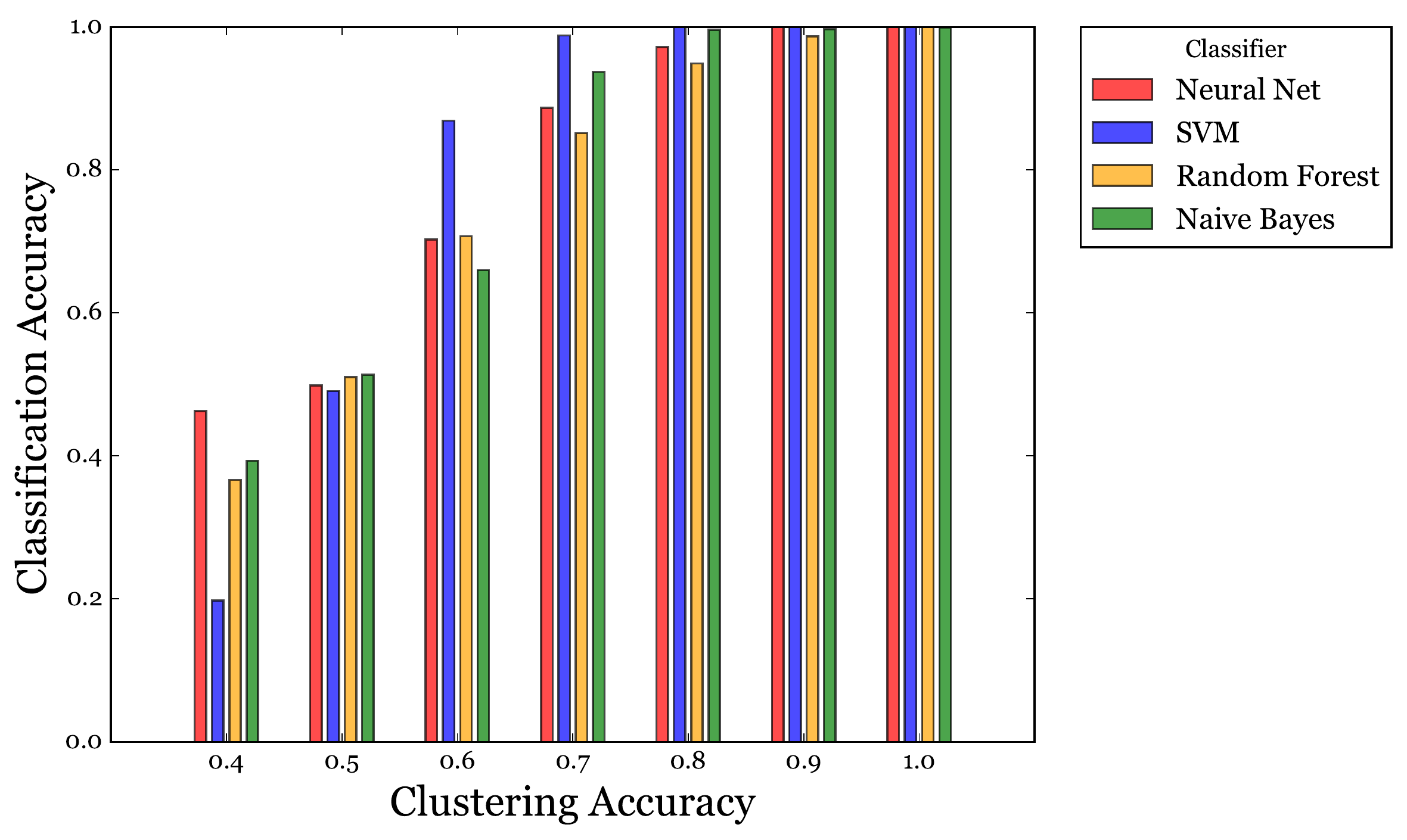}
    \vspace{-0.6cm}%
    \caption*{\footnotesize (b) (\emph{causative}) different clustering models accuracy based on non-adversarial data training.}
  \end{minipage}
  \caption{Accuracy of NB, RF, SVM and NN classifiers for varying adversarial strategies.}
    \label{fig_accuracy}
    \vspace{-0.4cm}
\end{figure*}

\emph{\textbf{Impact of Adversaries}:}~In our pertinent, well-motivated attack scenario, to assess the impact of collaborative pollution attacks, we examine the trade-off between the detectability of adversarial reports and the distortion they inflict on the system's output. Adversarial report distributions that significantly differ from the actual one (i.e., based on legitimate users’ reports) can more effectively distort the system’s output. But, at the same time, they can be detected and sifted easier. On the other hand, as modelled in Sec.~\ref{sec:problem_statement}, we also consider adversaries that employ a ``\emph{drifting strategy}'' (similar to concept drift in machine learning) to masquerade their existence. Such \emph{exploratory} attacks~\cite{kantchelian2013approaches} occur when adversaries react to the deployed classifiers and attempt to evade detection by strategically crafting malicious samples that are close to the data distribution(s) followed by the legitimate users.

In this context, Fig.~\ref{fig_accuracy} (a) shows the classification accuracy when different adversarial strategies are employed (different mean values). Here, we have again leveraged the synthetic datasets generated from the emulated traffic sensing task $(\mu = 16, \sigma = 2)$. We also fix the number of malicious users to be 40$\%$. The bars depict the precision score for all employed classifiers \emph{taking also into consideration the impact that adversarial strategies have on the descriptive data models produced by the clustering phase} (various clustering precision as depicted in Fig.~\ref{fig_clustering}). As we can see the overall correctness of the system remains high ($\geq 65\%$ while NB demonstrates an accuracy close to 85$\%$) for all cases where the difference in the mean values of the two distributions is (at least) 20$\%$. Recall that this reflects the percentage of overlapping regions between the distributions, meaning that in this case adversarial samples significantly differ from the actual ones. On the other hand, in the case of a ``\emph{drifting attack}'' (difference in the mean values is less than 15$\%$), we can see that the classification accuracy drops close to 60$\%$ whereas for NB drops to less than 40$\%$.

\noindent \textbf{Remark$\mathbf{\#5}$}: Classification accuracy is similar for NNs, SVM and RF, whereas NB behaves drastically worse for highly overlapping distributions. This is perhaps due to its simplicity in modelling, i.e., it assumes statistical independence between the different features. In our case, the feature used is a probability mass function of the data, which clearly violates the statistical independence assumption.

To alleviate for the above assumption on the correctness of the clustering accuracy, we also consider \emph{causative} attacks~\cite{kantchelian2013approaches} where adversaries try to influence the clustering process (training phase) in an attempt to reduce the quality of a future classifier. Due to the iterative nature of the data verification, exploratory attacks may ultimately have causative effects. This occurs when a drifting attack, mislabelled by the system, results to the invocation of the training phase leveraging this non-adversarial data model as training input data. Essentially this is an attack against the clustering process trying to create a false perception of the sensed phenomenon. The classifiers should react even in this case; its output must reflect the innate uncertainty of the sensed phenomena.

Fig.~\ref{fig_accuracy} (b) shows the accuracy of all employed classifiers for different clustering accuracies. This experiment reflects the impact that a ``\emph{drifting attack}'' will have on the re-clustering process (of the data verification) that will eventually affect the classification of subsequent user reports to legitimate and malicious. Again, the fraction of deviating users is set to 40$\%$. As we can see, when the clustering accuracy is $\geq 70\%$, all classifiers yield correct predictions with really high accuracy ($\geq 85\%$). What is interesting to notice is the generally better performance of the SVM: its accuracy increases rapidly  with respect to the other classifying algorithms. This can be attributed to its robustness in finding the optimal margin gap between separating hyper planes between the classes, especially when the quality of the training data gets better.

\noindent \textbf{Remark$\mathbf{\#6}$}: As expected, clustering accuracy can have a significant impact on the classification of subsequent user reports which is aggravated based on the vast amount of incoming data in IoT applications. As described in Sec.~\ref{sec:mcs}, the re-clustering process is invoked when a concept drift (which can also be the result of a drifting attack) has been detected which necessitates for the correct configuration of the \emph{sensitivity} parameter, $\theta$, so that the results of such attacks can be mitigated without depleting the system resources (which in turn can lead to denial of service attacks).

\section{Road-Map \& Future Prospects}
\label{sec:discussion}

As it is commonly the case for any relatively young research area, the landscape of Mobile Crowd-Sensing for IoT is fragmented into various families based on the emerging research challenges.~Undoubtedly, data trustworthiness is a prominent challenge with unprecedented number of consequences, should it is not addressed appropriately. We consider this paper as the first step towards the development of a holistic framework, which will improve data trustworthiness in IoT environments that utilize machine learning capabilities. Although, standard classification algorithms are not designed with such requirement in mind, in this paper we assessed their accuracy in presence of data infected with adversarial samples. We strongly believe that this work can be the basis of future research that will attempt to address two main challenges within the IoT security and privacy field: (a) \textit{accuracy}, in the context of concept drift, of IoT data and how this can be balanced with \textit{computational complexity}; and (b) \textit{near real-time performance}~of any proposed data trustworthiness framework in the presence of vast volume of IoT data processed; e.g., in the cloud in the form of big data or at the edge of a network 

Speaking about improving data verification, future work in the field can be geared towards proposing a combination of machine learning techniques to enhance classification accuracy within the investigated model. One shall use \textit{ensemble learning} to utilize multiple classifiers so that they can leverage their advantages and enhance the overall accuracy of IoT data verification \cite{wozniak2014survey}. However, ensemble learning will introduce high computational complexity. This generates by default an interesting challenge of investigating trade-offs between accuracy and complexity to determine optimal choices. We envisage that ensemble learning can alleviate the effects of concept drift, which refers to changes in the data distribution over time.
Another direction to improve IoT data trustworthiness is the application of deep learning by using autoencoders to train the deep neural network \cite{ngiam2011multimodal}. Due to this technique being computationally expensive, training must be done offline so that classification can be done online.

In the presence of vast volume of IoT data, it is often the case that limited computational resources (e.g., memory, time) and the requirement to make near real-time predictions affects the efficacy of various IoT applications: one of them being Mobile Crowd-Sensing (MCS). Furthermore, vast amount data may be collected so quickly that labelling all items may be delayed or even not possible.
To address these issues we envisage the use of game theory, which can determine optimal defence strategies, in the form of thresholds that determine when the system shall conduct certain required actions, such as re-clustering. Game theory can also support decisions of the defender, i.e., the IoT ecosystem itself, in presence of strategic attackers. These strategies will aim to maximize data trustworthiness in the presence of advanced colluding adversaries who shall utilize sophisticated adversarial strategies targeting data distortion by taking into account more parameters than our current model, such as: geography, users' density and number of submissions per second. A potential approach will seek optimal allocation of defending resources in a similar fashion to our previous work \cite{rontidis2015game}.

\section{Conclusions}
\label{sec:conclusion}

This paper extensively evaluates the effectiveness and accuracy of several fuzzy techniques within the IoT environment for Mobile Crowd-Sensing (MCS) applications. To the best of our knowledge, this is the first attempt at analyzing such data mining techniques in the context of secure and privacy-preserving MCS, where data trustworthiness is of paramount importance. Our work provides a comprehensive analysis on the applicability of various clustering and classification techniques for identifying descriptive and predictive models to analyze and predict the class of objects whose class label is unknown. We have evaluated the impact to the system output in the presence of strong colluding adversaries, and we have demonstrated how the spatio-temporal changes on the underlying phenomenon (i.e., concept drift) can masquerade the existence of attackers and affect the accuracy of clustering and classification processes. 
After this preliminary analysis, our future plans include exploiting more advanced techniques, including deep learning, so as to be able to propose a holistic framework that will improve IoT security and privacy using the right combination of machine learning models. This is a particularly challenging space due to the uncertainty of the classifiers with regards to the real nature of the reports submitted to a reporting station as legitimate.


\bibliographystyle{IEEEtran}
\bibliography{references}
\end{document}